\documentclass[letterpaper,twocolumn,10pt]{article}
\usepackage{usenix}

\usepackage{ising_v1}
\usepackage[normalem]{ulem}
\usepackage{tikz}
\usepackage{multirow}
\usepackage{diagbox}
\usepackage{algorithm}
\usepackage{algorithmic}
\usepackage[flushleft]{threeparttable}
\usepackage{subfig}
\usepackage{comment}
\usepackage{cite}
\usepackage{amsmath,amssymb,amsfonts}
\usepackage{graphicx}
\usepackage{textcomp}
\usepackage{xcolor}
\usepackage{fontawesome}

\def\submission{1}

\newcommand{\sys}{\texttt{FLASH-FHE}\xspace}

\title{\sys \text{\faBolt}: A Heterogeneous Architecture for Fully Homomorphic Encryption Acceleration}

\author{\rm
Junxue~Zhang$^1$,
~Xiaodian~Cheng$^2$\thanks{Work done while at iSINGLab @ Hong Kong University of Science and Technology.},
~Gang~Cao$^3$,
~Meng~Dai$^3$,
~Yijun~Sun$^1$,
~Han~Tian$^4$,\\
\rm
~Dian~Shen$^5$,
~Yong~Wang$^1$,
~Kai~Chen$^1$\\
$^1$iSINGLab @ Hong Kong University of Science and Technology\space\space
$^2$University of Waterloo\space\space
$^3$Clustar\\
$^4$University of Science and Technology of China\space\space
$^5$Southeast University\\
}

\begin{document}
\maketitle

\begin{abstract}
While many hardware accelerators have recently been proposed to address the inefficiency problem of fully homomorphic encryption~(FHE) schemes, none of them is able to deliver optimal performance when facing real-world FHE workloads consisting of a mixture of shallow and deep computations, due primarily to their homogeneous design principle.

This paper presents \sys, the first FHE accelerator with a heterogeneous architecture for mixed workloads. At its heart, \sys designs two types of computation clusters, \ie, bootstrappable and swift, to optimize for deep and shallow workloads respectively in terms of cryptographic parameters and hardware pipelines. We organize one bootstrappable and two swift clusters into one cluster affiliation, and present a scheduling scheme that provides sufficient acceleration for deep FHE workloads by utilizing all the affiliations, while improving parallelism for shallow FHE workloads by assigning one shallow workload per affiliation and dynamically decomposing the bootstrappable cluster into multiple swift pipelines to accelerate the assigned workload. We further show that these two types of clusters can share valuable on-chip memory, improving performance without significant resource consumption. We implement \sys with RTL and synthesize it using both 7nm and 14/12nm technology nodes, and our experiment results demonstrate that \sys achieves an average performance improvement of $1.4\times$ and $11.2\times$ compared to state-of-the-art FHE accelerators CraterLake and F1 for deep workloads, while delivering up to $8.0\times$ speedup for shallow workloads due to its heterogeneous architecture.
\end{abstract}
\section{Introduction}
\label{sec:introduction}

Fully Homomorphic Encryption~(FHE) has emerged as a promising technology for enabling privacy-preserving computation. It allows rich computation directly over encrypted data without the need for decryption. However, FHE suffers from a significant drawback—inefficiency. Compared to cleartext computation, FHE is approximately $5-6$ orders of magnitude slower. To address this challenge and make FHE a practical solution, researchers have proposed FHE accelerators.

FHE accelerators employ increasingly advanced hardware platforms, ranging from  CPUs~\cite{hexl} to GPUs~\cite{100x,tensorfhe}, FPGAs~\cite{hexl-fpga-url,heax,fab}, and ASICs~\cite{f1,craterlake,bts,ark,sharp}, in pursuit of high acceleration performance. However, in this paper, we observe that existing FHE accelerators all ignore one crucial feature of real-world FHE workloads: they are highly mixed, thus failing to achieve optimal overall performance for mixed workloads.

The mixed FHE workloads pose different design goals and choices for FHE accelerators. Specifically, some FHE workloads, referred to as shallow workloads, do not require bootstrapping, such as matrix computation and database queries~\cite{lookup-example-url}. They require a relatively small multiplication level $L$ and polynomial degree $N$. Therefore, a potential acceleration method for shallow workloads is to provide adequate parallelism for them. On the other hand, deep workloads, including neural network inference~\cite{lstm-ckks,resnet-ckks} and big data analytics~\cite{spark-fhe}, necessitate bootstrapping, along with requiring a larger $N$ and $L$. Moreover, the overall performance of deep workloads is largely decided by the performance of bootstrapping, which usually requires one dedicated accelerator to be fully accelerated.

Consequently, using FHE accelerators designed for shallow workloads, such as F1, to support mixed FHE workloads results in either the inability to handle deep workloads or over a $10\times$ performance degradation due to na\"ive extension~(\eg, F1+ in \cite{craterlake}). Conversely, utilizing FHE accelerators built for deep workloads, such as CraterLake, 
results in inadequate parallelism for shallow workloads. Increasing parallelism by allocating more computation clusters leads to overwhelming chip area and poses commercial drawbacks for FHE accelerators.
Furthermore, the non-continuous nature of cryptographic parameters for shallow and deep workloads eliminates the possibility of achieving good average performance through parameter averaging.

A potential solution to handling both shallow and deep workloads is to use two distinct architectures. However, Using two separate architectures for shallow and deep workloads nearly doubles production costs, while relying on CPU/GPU/FPGA for shallow workloads results in poor performance and additional costs. This also introduces deployment challenges, such as increased PCIe usage and failure rates.

\highlight{To this end, we ask:} \emph{can we design a FHE accelerator that achieves consistently high performance for mixed workloads?} To answer the question, we observe that existing FHE accelerators adopt a {\em homogeneous} design, where the cryptographic parameter, hardware pipeline design and scheduling mechanism are purely optimized for either shallow or deep workloads only, precluding simultaneous optimization for both shallow and deep workloads.

Based on this observation, we propose \sys, the FHE accelerator with a \emph{heterogeneous} architecture for mixed FHE workloads. At its core, \sys employs two types of computation clusters: bootstrappable and swift computation clusters, optimized for deep and shallow workloads, respectively. \sys organizes one bootstrappable and two swift clusters as one cluster affiliation and has eight affiliations in total. Besides conventional scheduling policy for deep workloads by using all affiliations, \sys further incorporates a scheduling mechanism that targets at improving parallelism by assigning one shallow workload to only one cluster affiliation, where the bootstrappable cluster is dynamically decomposed into multiple separate swift pipelines to adequately accelerate the assigned workload. Last but not least, \sys integrates a hierarchical data cache, which allows different clusters to share valuable on-chip memory resources. Therefore, we can improve the overall performance of \sys without dramatically increasing the chip area. To the best of our knowledge, \sys is among the first to introduce a heterogeneous architecture for FHE accelerators.

We implement \sys using RTL and synthesize it with two technology nodes, \ie, 7nm and 14/12nm, respectively. We further evaluate \sys with seven real-world FHE workloads, comprising three shallow and four deep workloads. 
Evaluation results demonstrate that compared to CraterLake and F1, which are two representative FHE accelerators with 14/12nm technology node, \sys can achieve $1.4\times$ and $11.2\times$ average performance improvement for deep workloads, respectively.
For shallow workloads, \sys can achieve up to $8.0\times$ speedup since it can accelerate multiple shallow workloads in parallel. Moreover, the significant performance improvement for shallow workloads is achieved at the cost of $<7\%$ extra hardware resources.

Finally, we summarize the contributions of this paper as follows:
\begin{icompact}
\item We thoroughly analyze representative shallow and deep FHE workloads and identify their optimal parameter settings, highlighting the polarization between these workloads.
\item Based on our observations, we are among the first to introduce a heterogeneous acceleration architecture for FHE, inspired by the big.LITTLE design philosophy. This includes heterogeneous (i)NTT pipelines, a multi-level transpose module, and a hierarchical caching structure to fully implement the big.LITTLE concept.
\item Leveraging our heterogeneous architecture, we propose the first multi-job scheduler for FHE workloads, optimizing both parallelization and cache hit ratios.
\end{icompact}

We believe our contributions provide valuable insights for the FHE community by demonstrating how the concept of heterogeneity can be seamlessly integrated into existing FHE accelerators, such as CraterLake~\cite{craterlake}, ARK~\cite{ark}, SHARP~\cite{sharp}, \etc. While these accelerators are primarily optimized for deep FHE workloads, our approach enables them to retain their current deep computation engines while incorporating key components such as multi-exit (i)NTT pipelines, multi-level transpose, and hierarchical caching. With the addition of these hardware modules and our multi-job scheduler, these accelerators can significantly improve their support for shallow FHE workloads by allowing parallel execution, making them more versatile and capable of handling mixed FHE workloads with minimal modifications.
\section{Background}
\label{sec:background}

\subsection{Fully Homomorphic Encryption}
\label{sec:background_fhe}
\begin{table}[t]
\scriptsize
\renewcommand\arraystretch{1.2}
\centering
\begin{tabularx}{\linewidth}{l l l}
\toprule
\bf Notation & \bf Definition \\
\midrule
$N$ & Degree of a polynomial. \\
$L$ & Multiplicative depth of a fresh ciphertext.  \\
$l$ & Current multiplicative depth of a ciphertext.  \\
$q$ & Coefficient modulus of cleartext polynomial. \\
$Q$ & Coefficient modulus of ciphertext polynomial. \\
$G$ & The number of computation groups in the group architecture. \\
$R$ & The number of rows in the NTT implementation. \\
$C$ & The number of columns in the NTT implementation. \\
\multirow{2}*{$l_{{\rm sub}}$} & The number of parallel modular multiplication pipelines \\
& in the basis converter of sequential BConv unit. \\
$\{q_i, i\in[0,L]\}$ & A set of moduli. $Q=\prod_{i=0}^Lq_i$. \\
$P$ & Special modulus for the keys. \\
$\{p_i,i\in[0,\alpha-1]\}$ & A set of special moduli. $P=\prod_{i=0}^{\alpha-1} p_i$. \\
${\rm dnum}$ & Decomposition number in key-switching. \\
$\alpha$ & $\#$ of special moduli $p_i$. $\alpha=\lfloor (L+1)/{\rm dnum}\rfloor $. \\
% $\{Q_j,j\in[0,{\rm dnum}]\} $ & A set of modulus factors. $Q_j=\prod_{i = j\alpha}^{(j+1)\alpha-1}q_i$. \\
% % $\mathbb{Z}_q[X]/(\Phi_M(X))$ & Plaintext space. A $M$-th cyclotomic polynomial ring. $M=2n$ \\
% % $\mathbb{Z}_Q^2[X]/(\Phi_M(X))$ Ciphertext space.  & \\
% $\Delta$ & Scaling factor. \\
% $m$ & Plaintext. \\
% $s$ & Secret key. \\
% $\boldsymbol{\bf pk}$ & Public key. \\
% $\boldsymbol{\bf ct}$ & Ciphertext. \\
% $\boldsymbol{\bf swk}$ & Switching key for key-switching. \\
% $\kappa_k$ & \multirow{2}{*}{\shortstack[l]{An automorphism on the polynomial which \\ permutes the position of the coefficients.}} \\
% & \\
% Dyadic multiplication & Element-wise multiplication of polynomials.\\
\bottomrule
\end{tabularx}
\caption{Notations used in this paper.}
\label{tab:notations}
\end{table}

FHE allows performing specific homomorphic operations directly over ciphertexts without decrypting them~\cite{ckks,bfv,bgv,tfhe}. In the following sections, we will take CKKS~\cite{ckks} as an example to demonstrate the operations used in FHE, and other FHE schemes, such as BFV~\cite{bfv}, BGV~\cite{bgv}, \etc, should share similar operations. Table~\ref{tab:notations} summarizes the notations used in this paper.

\parab{Homomorphic Addition:} The ciphertext in CKKS can be represented as ${\boldsymbol{\rm ct}}=(c_0,c_1)$, where $c$ is a polynomial. The result of the homomorphic addition between one ciphertext ${\boldsymbol{\rm ct}}$ and one cleartext $m$, which is also a polynomial, is $(c_0+m,c_1)$ while the result of two ciphertexts ${\boldsymbol{\rm ct_0}}=(c_{0,0},c_{0,1})$ and ${\boldsymbol{\rm ct_1}}=(c_{1,0},c_{1,1})$ is $(c_{0,0}+c_{1,0},c_{0,1}+c_{1,1})$. The addition operator $+$ here denotes an element-wise addition of two polynomials.

\parab{Homomorphic Multiplication:} The multiplication result between one ciphertext ${\boldsymbol{\rm ct}}$ and one cleartext $m$ is $(c_0*m,c_1*m)$, where $*$ represents polynomial multiplications. However, the homomorphic multiplication operation between two ciphertexts is more complicated since multiplication between the above $\boldsymbol{\rm ct_0}$ and $\boldsymbol{\rm ct_1}$ results in a ciphertext $\boldsymbol{\rm ct'}$ with three elements:

\vspace{-4ex}
\begin{equation}
\small
\boldsymbol{\rm ct'}
%=(d_0,d_1,d_2)
=(c_{0,0} * c_{1,0}, c_{0,0} * c_{1,1}+ c_{0,1} * c_{1,0}, c_{0,1} * c_{1,1})
\end{equation}
\vspace{-4ex}

To turn the ciphertext back into two elements for future ciphertext computation with consistent form, we have to perform ciphertext maintenance operation -- key-switching -- to relinearize the ciphertext. We will introduce the key-switching operation later.

\parab{Rotation:} CKKS exploits the idea of SIMD~(\ie, batching) by packing multiple cleartexts in a single vector, which is further encrypted as a single ciphertext. Although it is efficient for most operations, it requires homomorphic ciphertext rotation to perform intra-batch calculations. The ciphertext rotation consist of two operations. First, it performs automorphisms operation, which is a permutation over coefficients of polynomials for both ciphertext $\boldsymbol{\rm ct}$ and secret key $s$. Second, key-switching is performed to convert the permutated ciphertext into the final result.

Besides the aforementioned homomorphic evaluation operations, FHE schemes have to perform some ciphertext maintenance operations to preserve the correctness of these operations. Two of the most important operations are key-switching and bootstrapping.

\parab{Key-switching:} As its name implies, the key-switching operation homomorphically switches the secret key of a ciphertext while keeping the corresponding cleartext unchanged. Therefore, the key-switching operation is extensively used in homomorphic operations, such as multiplication, rotation, \etc, to keep the results correct. The key-switching operation accounts for a significant portion of the overall computation time due to the inherent complexity involved in its internal computations.

\parab{Bootstrapping:} After performing homomorphic operations, especially the multiplication operation, the noise of a ciphertext increase. When too many operations are performed, the noise will overflow, leading to incorrect results after decryption. We call the maximum number of consecutive homomorphic multiplications allowed as the multiplicative level/budget. Therefore, before the number of operations exceeds the multiplicative level, the bootstrapping operation is needed to "refresh" the noise of ciphertexts by homomorphically re-encrypting the old ciphertexts to generate a fresh one.

Bootstrapping is the most complicated operation in FHE, which involves many other operations including key-switching. If used, bootstrapping will considerably determine the performance of the overall performance of a FHE scheme.

Finally, we discuss the cryptographic parameter choices of FHE schemes. There are several important parameters to determine: (1) the polynomial degree $N$; (2) the range of the coefficients of a the polynomial. Since CKKS is built on a polynomial ring, the range of the coefficients is decided by the modulus $Q$. 3) the modulus $P$ used in keys. To ensure security, $N/\log PQ$ must be over a certain threshold, for example, $128$ bits~\cite{fhe-security}. Moreover, $Q$ determines the multiplicative level/budget, and a larger $Q$ generally leads to more multiplicative levels. For example, multiplicative level of $16$ requires $Q$ to be around $512$. Therefore, to satisfy both high security and high multiplicative levels, a larger $N$ is also needed, which inevitably causes inflated ciphertexts.

\subsection{General Optimizations for FHE Implementations}
\label{sec:background_fhe_optimization}

This section introduces three general optimization techniques used by modern FHE implementations.

\parab{NTT/iNTT:} To accelerate polynomial multiplications, the Number Theoretic Transform~(NTT) and the inverse Number Theoretic Transform~(iNTT) operations are used. The NTT/iNTT lowers the time complexity of polynomial multiplications from $O(n^2)$ to $O(n\log n)$, where $n$ is the degree of a polynomial. Readers can regard NTT/iNTT as Fast Fourier Transform over finite field. The basic operation of NTT/iNTT is the butterfly operation~\cite{ct_butterfly,gs_butterfly}.

\parab{RNS Decomposition:} 
To reduce computation complexity in CKKS, large modulus $Q$ and $P$ are decomposed into the product of several smaller co-prime moduli: $Q=q_0q_1...q_L$ and $P=p_0p_1...p_{\alpha}$. A set containing multiple co-prime moduli $\{q_0, q_1, \cdots\}$ is called an RNS basis. Following the Chinese Reminder Theorem~\cite{crt}, a large integer can be represented by taking its modulus results with respect to all moduli in the RNS basis. This representation is referred to as the RNS representation of the integer.

\parab{Fast Basis Conversion (BConv)}: Some FHE operations involve the conversion~(switching) of modulus, which requires special operations under RNS representation. For example, the Fast Basis Conversion (BConv) approximates the modulus switching by converting an integer from the original RNS basis to a new basis. BConv mainly consists of multiple modular multiplications and independent summations.

\subsection{Prior FHE Accelerators}
\label{sec:background_fhe_acc}

One key problem restricting FHE from practical deployment is its inefficiency issue. Even highly optimized FHE schemes suffer from five orders of magnitudes slower performance than cleartexts computation~\cite{sok_fhe_acc}. To solve the problem, various FHE accelerators are proposed~\cite{hexl-fpga-url,heax,fab,100x,tensorfhe,f1,craterlake,bts,ark}. We categorize them based on the employed hardware platform.

\parab{FPGA:} Some FHE accelerators leverage Field Programmable Gate Array~(FPGA) as their hardware platform, such as Intel HEXL-FPGA~\cite{hexl-fpga-url}, HEAX~\cite{heax}, FAB~\cite{fab}. However, due to the limitations of FPGA itself, \eg, limited programmable resources, low operational frequency and low memory bandwidth, FPGA-based FHE accelerators cannot satisfy the requirements of FHE schemes, which are both computation and memory intensive. Therefore, their performance is far from optimal.

\parab{GPGPU:} General Purpose GPU~(GPGPU) has been extensively adopted to offload AI-related applications. Due to their relatively good performance to accelerate data-parallel applications, researchers also utilize GPGPUs to accelerate FHE applications, such as $100\times$~\cite{100x}, TensorFHE~\cite{tensorfhe}, \etc. However, as revealed in~\cite{flash}, GPGPU suffers from architectural deficiency when handling inflated data, causing impractical performance. Moreover, large power consumption of GPGPU is another drawback of these solutions.

\parab{ASIC:} Recently, people have begun to explore Application Specific Integrated Circuit~(ASIC) for superb performance. F1 is the first programmable FHE accelerator implemented with ASIC~\cite{f1}, which can outperform software FHE implementations by $5400\times$ on average.
However, due to the lack of efficient bootstrapping, F1 can only deliver ideal acceleration for shallow FHE workloads.
Later, CraterLake~\cite{craterlake} and BTS~\cite{bts} were proposed to support full bootstrapping and both employ massive computation units, such as NTT/iNTT, and considerable amounts of on-chip memory.
ARK~\cite{ark} is the successor of them and it further improves the bootstrapping performance by an algorithm-architecture co-design.

\begin{figure*}[t]
\centering
\subfloat[][Matrix Multiplication]{
\includegraphics[width=0.23\textwidth]{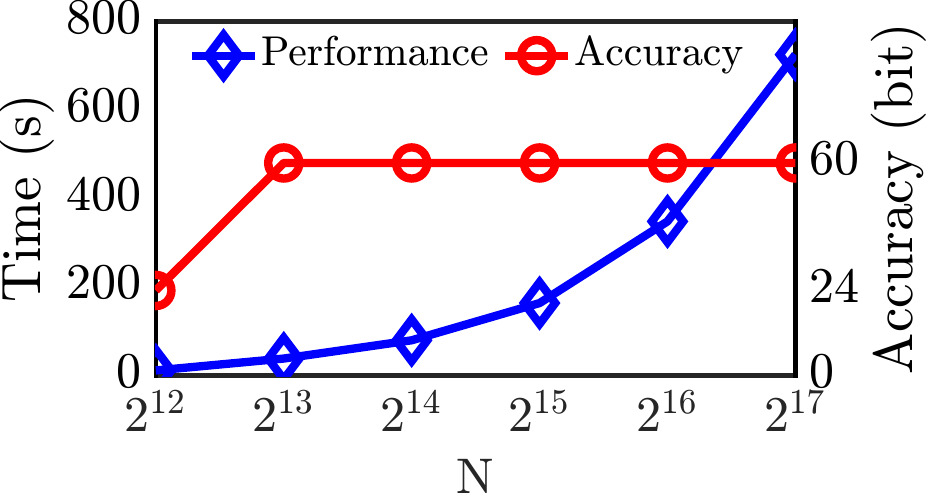}
\label{fig:motivation-performance-matrix-mul}
}
\hfill
\subfloat[][DBTable Lookup]{
\includegraphics[width=0.23\textwidth]{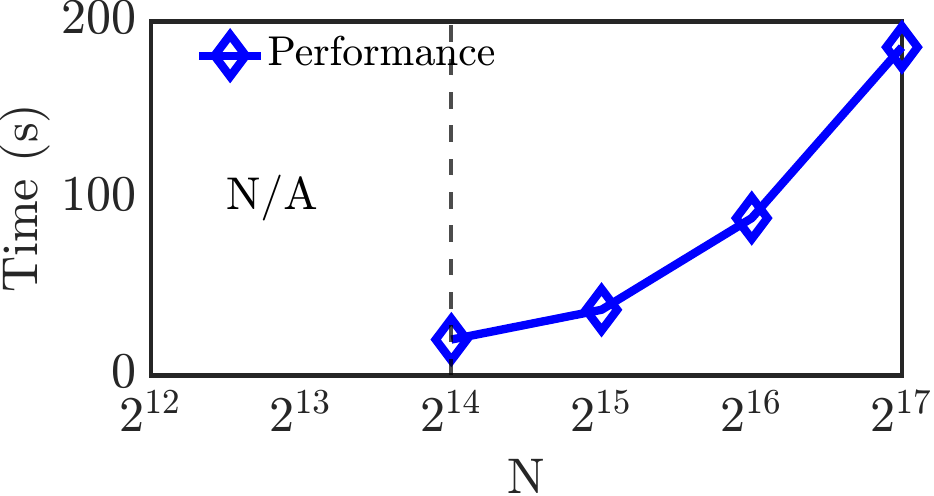}
\label{fig:motivation-performance-table-lookup}
}
\hfill
\subfloat[][Logistic Regression]{
\includegraphics[width=0.23\textwidth]{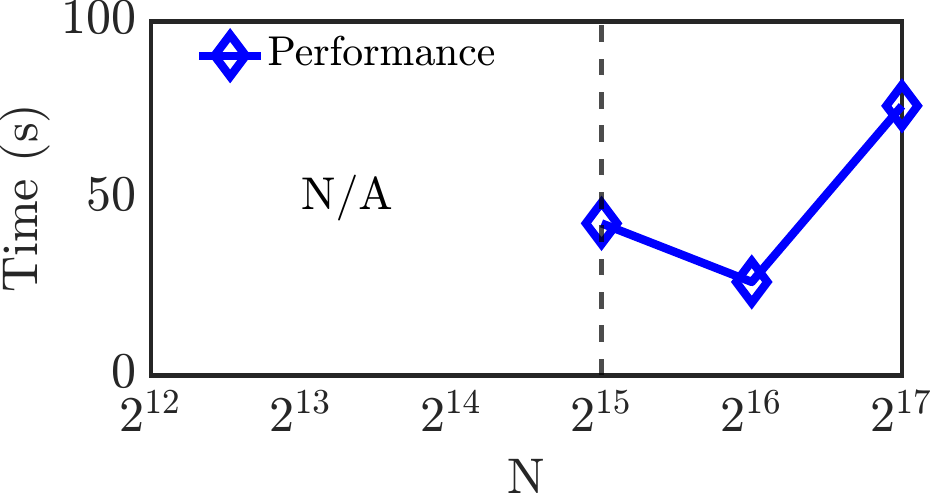}
\label{fig:motivation-performance-helr}
}
\hfill
\subfloat[][LSTM]{
\includegraphics[width=0.23\textwidth]{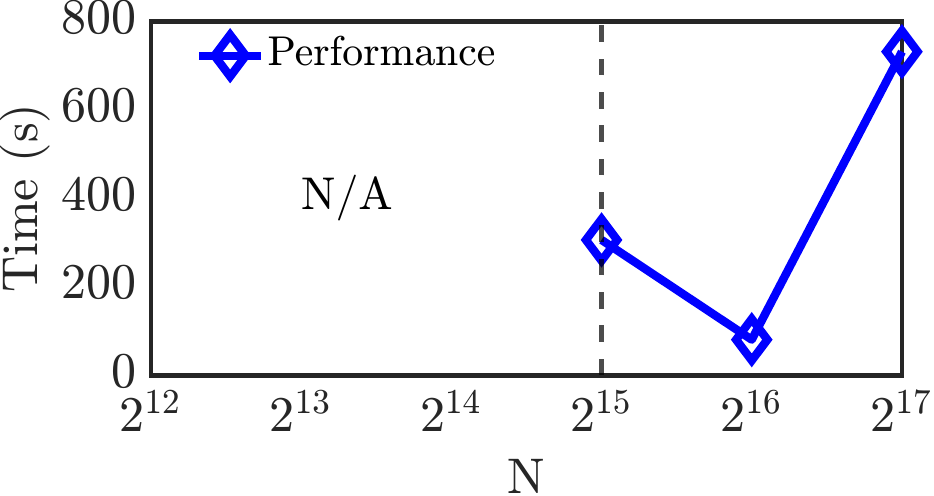}
\label{fig:motivation-performance-lstm}
}
\caption{Performance of FHE Workloads.}
\label{fig:motivation-performance}
\end{figure*} 

Existing development trend of FHE accelerators is to support bootstrapping, thus mainly targeting improving the performance of deep FHE workloads. However, in this paper, we argue that real-world FHE workloads are highly mixed of shallow and deep ones, thus significant opportunities exist to enhance the performance of existing FHE accelerators by considering how to simultaneously accelerate both workloads.
\section{Motivation}
\label{sec:analysis}

In this section, we first introduce the nature of FHE workloads that they are highly mixed --- some of the workloads require bootstrapping while some do not~(\secref{sec:analysis_workloads}). Second, we will demonstrate how different cryptographic parameters impact the end-to-end performance, and present the appropriate parameter choices for both workloads~(\secref{sec:analysis_crypto_params}). 
Third, we show that both hardware pipeline design and scheduling policies are completely different for both workloads~(\secref{sec:impact_pipeline_design}).

\subsection{Mixed FHE Workloads}
\label{sec:analysis_workloads}

FHE workloads exhibit a high degree of variability in real-world scenarios, encompassing both bootstrapping and non-bootstrapping requirements. In this paper, similar to \cite{craterlake}, we call workloads that require bootstrapping as deep workloads and those without bootstrapping as shallow workloads.

Shallow FHE workloads are common in real-world. First, some FHE workloads do not require very deep multiplications by themselves, such as database query~\cite{lookup-example-url}, private information retrieval~\cite{seal-pir}, Beaver's multiplication triples generation for secret sharing~\cite{mtg-rlwe}, federated learning~\cite{fl-fhe}, shallow neural network inference~\cite{lola}, \etc. Second, people also try to avoid bootstrapping by introducing some stochastic designs for even more complicated applications to preserve acceptable performance. For example, Sphinx synthesizes FHE and differential privacy to achieve a uniformed framework for neural network training and inference without bootstrapping~(but requires decryption and re-cryption at client side)~\cite{sphinx}.

With the increasing requirements of applying privacy-preserving technologies to more complicated applications, deep FHE workloads also become widely adopted. One typical example is machine learning training based on pure FHE, such as FHE-protected logistic regression training~\cite{helr}, FHE-protected deep neural networking training~\cite{nn-training-fhe}, \etc. 

FHE accelerators, as an essential infrastructure, should deliver optimal acceleration for both workloads simultaneously. 
However, in the following sections, we will demonstrate that these workloads require different choices of cryptographic parameters~(\secref{sec:analysis_crypto_params}), pipeline designs and scheduling policies~(\secref{sec:impact_pipeline_design}), thus causing sub-optimal performance for existing FHE accelerators which utilize a homogeneous architecture.

\subsection{Impact of Cryptographic Parameters}
\label{sec:analysis_crypto_params}

\begin{table}[t]
\scriptsize
\centering
\begin{tabularx}{\linewidth}{l X X X X X X}
\toprule
& $2^{12}$ & $2^{13}$ & $2^{14}$ & $2^{15}$ & $2^{16}$ & $2^{17}$ \\
\midrule
Matrix Multiplication & 101 & 192 & 192 & 192 & 192 & 192 \\
DBTable Lookup & - & - & 399 & 575 & 653 & 653 \\
Logistic Regression & - & - & - & 816 & 1550 & 3125 \\
LSTM & - & - & - & 821 & 1536 & 2901 \\
\bottomrule
\end{tabularx}
\caption{The $\log PQ$ setting used in the paper.}
\label{tab:backgroud-workloads}
\end{table}

First, we qualitatively introduce the relationship of some important parameters. As discussed in \secref{sec:background_fhe}, for deep FHE workloads, to support bootstrapping, a large multiplication depth $L$ is needed since (1) frequent bootstrapping should be avoided and (2) bootstrapping itself consumes massive multiplication depths. Moreover, when a large $L$ is chosen, a large polynomial degree $N$ is also required to ensure security~\cite{fhe-security}. In contrast, for shallow workloads, the $N$ and $L$ are usually much smaller.
Next, we present how these parameters impact the performance of both shallow and deep workloads through testbed experiments. Especially, we implement the following four real-world workloads with \texttt{Lattigo}~\cite{lattigo-url}, a widely-adopted FHE library.

\begin{icompact}
\item \textbf{Matrix Multiplication:} It's a shallow FHE workloads. We multiply two matrix of $100\times 1000$ and $1000\times10$ elements encrypted via CKKS, respectively.
\item \textbf{DBTable Lookup:} It's a shallow FHE workloads. We refer the implementation used in \cite{lookup-example-url} and use BGV to encrypt the data. We have modified the algorithm to use binary encoding to encode the key to reduce the required multiplication levels.
\item \textbf{Logistic Regression:} It's a deep FHE workload and the implementation is based on \cite{helr}. We measure the training time of a single batch with up to 197 features and 50 samples per batch. We perform bootstrapping once the multiplication level is exhausted.
\item \textbf{LSTM:} Long Short-Term Memory~(LSTM) is a recurrent neural network~(RNN), which is widely adopted to learn the long-term dependencies, especially in sequence prediction problems. In this paper, we follow the implementation in \cite{lstm-ckks} to use CKKS to protected the parameters of a LSTM network. It's also a deep FHE workload.
\end{icompact}

In every workload, we fix the security level to be $128$bit and vary the $N$ from $2^{12}$ to $2^{17}$. Therefore, to ensure the $128$bit security level, the maximum $\log PQ$ should vary from $101$ to $3125$ accordingly. We will choose the most suitable multiplication depth $L$ for each workloads while not violating the constraints of maximum $\log PQ$. 
Another cryptographic parameter that affects the performance of bootstrapping is the decomposition number, $\rm dnum$, which is used in the key-switching operations. As suggested in some papers~\cite{100x,bts}, to support efficient bootstrapping for deep workloads, we usually choose a small $\rm dnum$. However, a small $\rm dnum$ may violate the security guarantee. Thus, we try to choose a small $\rm dnum$ that will not violate the security guarantee when $N$ and $L$ are fixed. The detailed settings of each workload are shown in \tabref{tab:backgroud-workloads}.

In the conducted experiments, we focused on measuring the execution time of the FHE workloads. Specifically, for the Logistic Regression workload, we calculated the average execution time of one iteration. As for the LSTM workload, we reported the execution time of one LSTM unit. The obtained results are illustrated in  \figref{fig:motivation-performance}. In summary, we have the following observations:

\begin{icompact}
\item Shallow workloads, such as Matrix Multiplication~(\figref{fig:motivation-performance-matrix-mul}) and DBTable Lookup~(\figref{fig:motivation-performance-table-lookup}), can achieve ideal performance with small $N$ and $L$. A large $N$ and $L$ in turn leads to performance degradation. For example, for Matrix Multiplication, when the $N$ equals $2^{13}$, the application can simultaneously achieve adequate accuracy~(60bit) while keeping efficient. For DBTable Lookup, when the $N$ equals $2^{14}$, it can achieve highly efficient performance. With analysis on more shallow workloads~(which are not shown due to limited space), we find $N \le 2^{14}$ is a reasonable value for shallow FHE workloads.
\item Deep workloads, such as Logistic Regression~(\figref{fig:motivation-performance-helr}) and LSTM~(\figref{fig:motivation-performance-lstm}), require a large $N$ and $L$ to work properly. For example, if the $N<2^{15}$, the Logistic Regression cannot even finish one iteration, leading to task failure. Moreover, when the $N$~(also $\log PQ$) becomes larger, the performance of the FHE application first improves, then drops. The reason is that if $N$ is small, the multiplicative depth is low. As a result, the bootstrapping is too frequent, degrading the overall performance. In contrast, if $N$ is too large, the performance also degrades due to the increased complexity of cryptographic system. In summary, we observe that $2^{15} \le N \le 2^{16}$ is a sweet range for most deep FHE workloads.
\end{icompact}

\begin{figure}[t]
\centering
\includegraphics[width=\columnwidth]{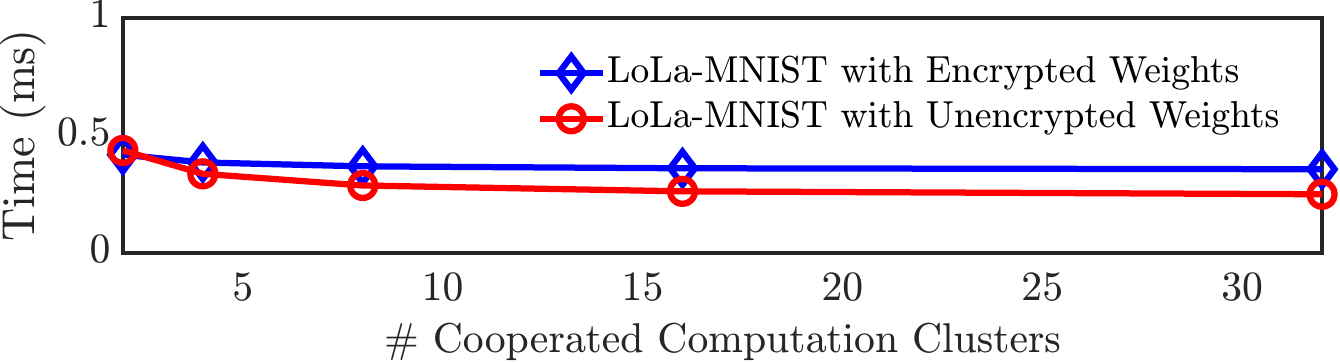}
\caption{The performance of shallow FHE workloads.}
\label{fig:motivation-shallow-ws-over-cores}
\end{figure}

Consequently, FHE accelerators designed for shallow workloads, such as F1~\cite{f1}, either fail to run deep workloads or suffer from enormous performance reduction. The reason is that, since they are optimized for shallow workloads, they usually choose to support a small $N$ and $L$. For example, F1 supports $N\leq 2^{14}$. Thus, deep workloads, such as Logistic Regression, cannot even run when the $N \leq 2^{14}$. Furthermore, as pointed out in the previous work~\cite{craterlake}, if we arbitrarily increase the $N$ of F1 to $2^{16}$~(denoted as F1+ in Table 3 of \cite{craterlake}), its performance is still more than $10\times$ slower than expected. The reason is that it lacks an efficient bootstrapping implementation~(details in \secref{sec:impact_pipeline_design}).

\subsection{Impact of Pipeline Design \& Scheduling}
\label{sec:impact_pipeline_design}
As discussed in previous literature~\cite{sok_fhe_acc}, bootstrapping occupies more than $80\%$ of the total computation time. Therefore, efficiently accelerating bootstrapping is a key to improving the performance of deep FHE workloads. 

Recent works have explored to build a dedicated iNTT $\rightarrow$ BConv $\rightarrow$ NTT pipeline to enable efficient bootstrapping computation for deep workloads~\cite{craterlake,bts,ark,sharp}. Moreover, the pipeline execution involves the cooperation of all computation clusters due to the heavy workload. Therefore, the scheduler of recent FHE accelerators tends to schedule the whole accelerator to expedite one dedicated deep FHE workload, and a development trend is to utilize more hardware resources for better performance~\cite{f1,craterlake,bts,ark,sharp}.

On the contrary, in this paper, we find that for shallow workloads, blindly allocating more hardware resources, \eg, computation clusters, does not lead to effective acceleration. To demonstrate this, we use experiments to investigate how the performance of two shallow FHE workloads changes when we only increase the number of computation clusters without changing the workloads. To align with most results in previous works, we use LoLa-MNIST with and without Unencrypted Weights as workloads here~(extensive details in \secref{sec:evaluation_methodology}). The results are presented in \figref{fig:motivation-shallow-ws-over-cores} and we can observe that when we allocate more than four 128-bit NTT computation clusters to cooperatively accelerate one shallow FHE workload, the performance improves very little. Instead, we argue that for shallow FHE workloads, the optimal scheduling policy is to distribute multiple shallow FHE workloads on one accelerator in parallel while concurrently implementing sufficient computation clusters. 

As a result, if we use existing FHE accelerators that are designed for deep FHE tasks to accelerate the shallow FHE workloads, they can only handle one job simultaneously, leading to low data parallelism. Meanwhile, na\"ively increasing the number of clusters causes overwhelming chip area consumption, which is impractical.

\parab{Conclusion:} Existing FHE accelerators cannot consistently achieve high performance facing real-world mixed FHE workloads. The crux lies in the fact that they all adopt a \emph{homogeneous} design: their target cryptographic parameters, pipeline design and scheduling policy are optimized for either shallow or deep FHE workload.

\subsection{Two Different Architectures?}
\label{sec:two_different_architecture}
A straightforward approach would be to utilize two distinct architectures for shallow and deep workloads, respectively. However, this solution either incurs significantly higher production costs or results in degraded performance, along with practical deployment challenges.
\begin{icompact}
\item \emph{Using two ASICs:} One option is to produce two separate ASICs, each dedicated to either shallow or deep workloads. However, this approach nearly doubles production costs due to the non-recurring engineering (NRE) costs—such as design, verification, and mask set creation—that are required for each chip.
\item \emph{Using CPU/GPU/FPGA to handle shallow workloads:} Another option is to use a dedicated ASIC for deep FHE workloads and rely on CPU/GPU/FPGA for shallow workloads. However, as demonstrated in \S\ref{sec:evaluation}, CPU/GPU/FPGA platforms fail to deliver competitive performance. Additionally, GPUs and FPGAs introduce further costs.
\end{icompact}
Furthermore, both of these solutions present deployment challenges, including increased PCIe slot usage, higher failure rates, \etc.
\section{\sys}
\label{sec:design}

\begin{figure}[t]
\centering
\includegraphics[width=\columnwidth]{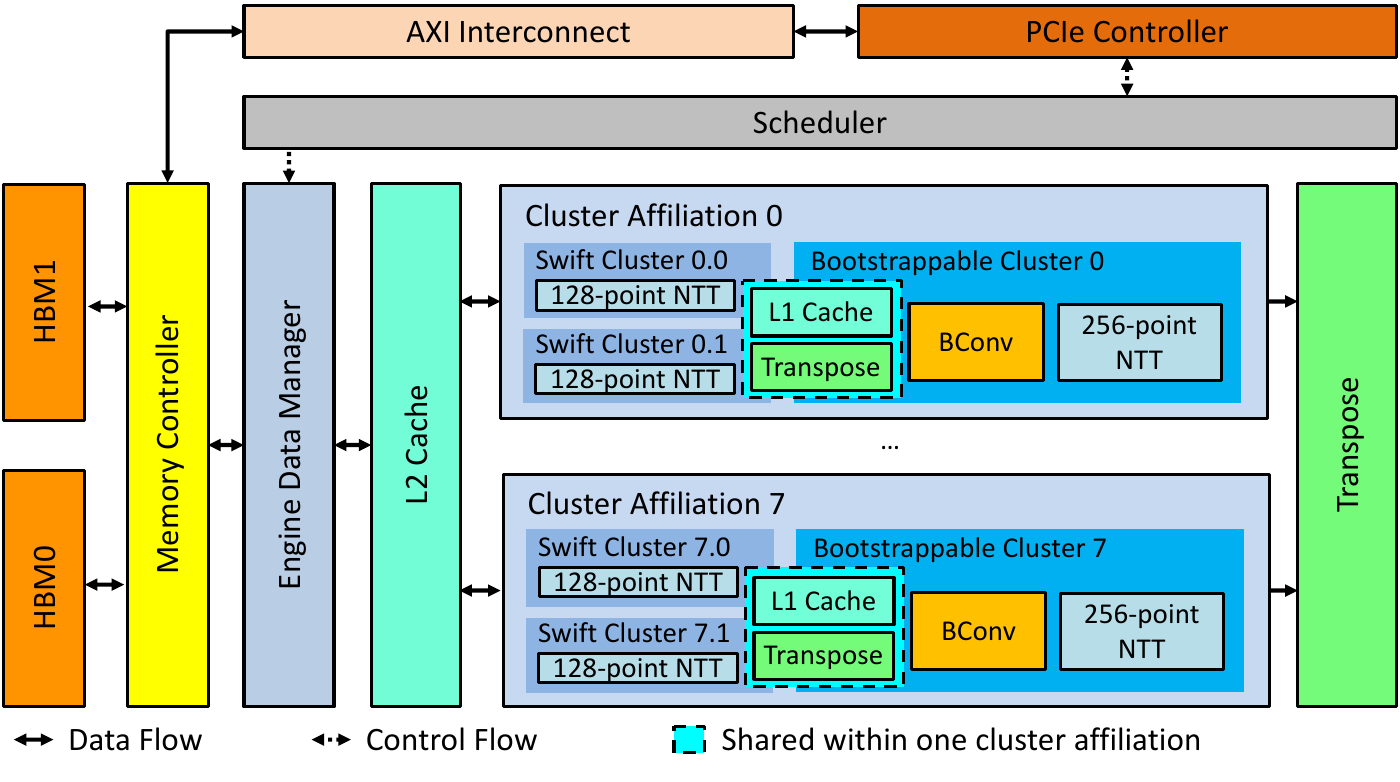}
\caption{\sys's architectural overview.}
\label{fig:design-arch-overview}
\end{figure}

To solve this problem, we propose \sys, a practical FHE accelerator with a \emph{heterogeneous} architecture. Instead of relying on homogeneous computation clusters, \sys incorporates two types of clusters: bootstrappable and swift. In terms of cryptographic parameters and pipeline design, these clusters are optimized for deep and shallow workloads, respectively. We organize one bootstrappable and two swift clusters as one cluster affiliation and \sys has eight affiliations in total~(\secref{sec:design_computation_cluster}).
Then, \sys's scheduler tries to maximize the parallelism for shallow FHE workloads by assigning one shallow workload to only one cluster affiliation, which can deliver effective acceleration by decomposing one bootstrappable cluster into multiple suitable pipelines for shallow workloads. In this way, \sys can use eight cluster affiliations to achieve a parallelism of eight~(\secref{sec:design_scheduler}).
Furthermore, we introduce the design of a hierarchical data cache that is shared between the bootstrappable and swift clusters. This approach allows us to enhance performance by allocating additional computation clusters without significantly increasing the demand for on-chip memory resources~(\secref{sec:design_cache}).

\parab{Architectural Overview:} \figref{fig:design-arch-overview} illustrates the architectural overview of \sys. It comprises three key components: computation clusters, control subsystem, and memory subsystem. As previously mentioned, the computation clusters consist of eight cluster affiliations, each housing one bootstrappable and two swift clusters. The clusters are connected using a multi-leveled transpose module. The core module of the control subsystem is the scheduler, and \sys also leverages PCIe and memory controller for auxiliary functions. In terms of the memory subsystem, \sys employs two high-bandwidth memory~(HBM)\cite{hbm-url} units for off-chip data storage, and incorporates a hierarchical data cache~(L1/L2) for on-chip storage.

\subsection{Computation Clusters}
\label{sec:design_computation_cluster}

In this section, we first introduce the design of the two types of computation clusters used in \sys, and then present how we use cluster affiliation to organize them.

\parab{Bootstrappable Clusters:} To ensure efficient bootstrapping, we implement a complete iNTT $\rightarrow$ BConv $\rightarrow$ NTT pipeline for bootstrappable clusters, following previous works~\cite{craterlake,bts,ark,sharp}. Within a bootstrappable cluster, we incorporate one (i)NTT pipeline and one BConv module.

\begin{figure}[t]
\includegraphics[width=\columnwidth]{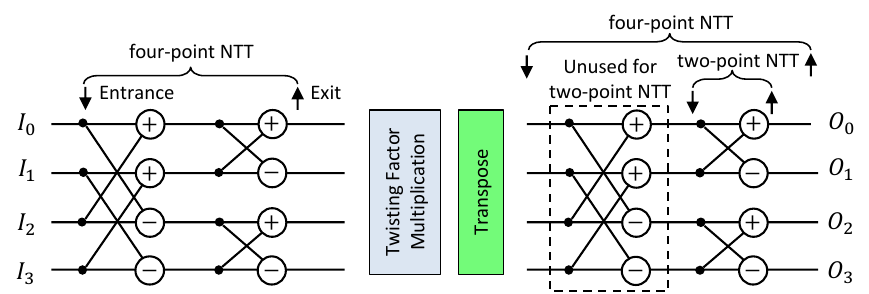}
\caption{(i)NTT workflow. The (i)NTT pipeline in \sys supports multiple entrances and multiple exits to enable efficient execution of varying-point NTTs.}
\label{fig:design-ntt-workflow}
\end{figure}

The (i)NTT pipeline consists of a four-step (i)NTT\footnote{The four-step NTT algorithm has been comprehensively introduced in previous works~\cite{f1,craterlake,sok_fhe_acc}, thus we omit the detailed algorithm in this paper.} module with $N=2^{16}$, encompassing a $\sqrt{N}$-point~(i.e., $2^8$-point) standard (i)NTT circuit, a multiplication circuit, and a transpose circuit~(L1, more details in the following sections). A simplified example of the NTT workflow for $N=2^4$ is depicted in \figref{fig:design-ntt-workflow}. The input data undergoes the standard (i)NTT circuit in the first step, followed by processing through the multiplication and transpose circuits. Finally, the transposed data is fed back into the (i)NTT circuit for a second round, resulting in the final results. To accommodate dynamic changes in the value of $N$, we extend the (i)NTT circuit to support multiple entry and exit points. This allows data to be injected or fetched at any position within the circuit, enabling the pipeline to support (i)NTT with $N \le 2^{16}$. One example is shown in the right part of \figref{fig:design-ntt-workflow}, we can use the 4-point NTT pipeline to support two parallel 2-point NTT computations. Therefore, our design enables the decomposition of the large (i)NTT circuit into multiple parallel smaller (i)NTT circuits, boosting the acceleration of workloads requiring a smaller NTT, \eg, shallow workload. As a result, \sys's bootstrappable computation clusters can support both deep FHE workloads and some parallel shallow workloads~(further details of scheduling policy will be discussed in \secref{sec:design_scheduler}).

\begin{figure}[t]
\centering
\includegraphics[width=\columnwidth]{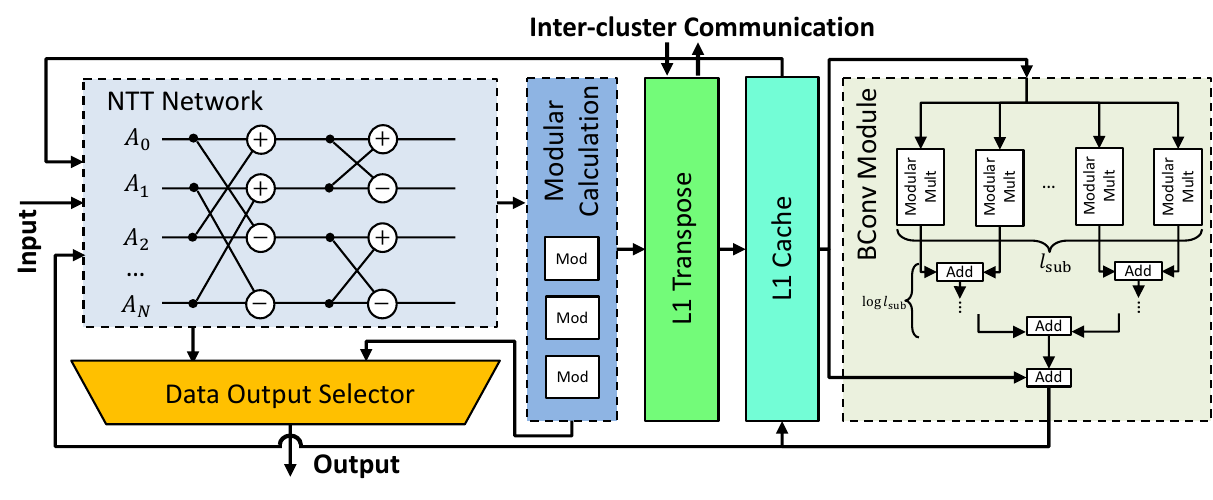}
\caption{The physical design of the bootstrappable computation cluster.}
\label{fig:design_bootstrappable_engine}
\end{figure}

\figref{fig:design_bootstrappable_engine} presents the physical design of the bootstrappable computation cluster. The left part demonstrates the above-introduced iNTT pipeline. The right part is the BConv module, which performs $l_{\rm sub}$ parallel modular multiplications and sums their results to obtain the converted RNS basis. We choose $l_{\rm sub}=60$ to maximize the performance of BConv in the key-switching pipeline. The two parts are connected via a L1-transpose module, which will be introduced later in this section. To accommodate various modes, all modules within the bootstrappable cluster can be bypassed. For instance, when accelerating shallow workloads, the BConv module can be entirely bypassed since shallow FHE workloads do not require the key-switching operation which involves the iNTT $\rightarrow$ BConv $\rightarrow$ NTT pipeline.

\parab{Swift Clusters:} The swift computation clusters are dedicated to shallow workloads. Therefore, the BConv module is not allocated within the swift clusters. Instead, we design one (i)NTT pipeline with $N=2^{14}$ in each swift cluster, as $N=2^{14}$ is sufficient for shallow workloads at the 128-bit security level. Similar to the bootstrappable clusters, the (i)NTT pipeline within the swift clusters comprises a $2^7$-point standard (i)NTT circuit, a multiplication circuit, and a transpose circuit.

\parab{Cluster Affiliation:} In the design of \sys, one bootstrappable cluster and two swift clusters are organized as a cluster affiliation, and we allocate eight cluster affiliations in total.
The choice of the one-to-two ratio is guided by both the \emph{parameter considerations for the clusters} and \emph{hardware design efficiency}. As previously mentioned, based on the analysis of representative workloads, the bootstrappable cluster is equipped with a $2^8$-point (i)NTT circuit, while the swift cluster contains a $2^7$-point (i)NTT circuit. This configuration allows for a straightforward and efficient connection of one $2^8$-point (i)NTT circuit with two $2^7$-point (i)NTT circuits using a transpose module (further details are provided in \textbf{Multi-level Transpose}). Additionally, the four $2^7$-point (i)NTT circuits within a cluster affiliation can effectively accelerate a single shallow FHE workload without requiring the involvement of other cluster affiliations.

It’s important to note that this ratio is not dictated by the proportion of deep to shallow FHE workloads in mixed workloads. Instead, \sys utilizes the scheduler to handle dynamic workloads~(more details in \S\ref{sec:design_scheduler}).

As a result, different from previous FHE accelerators that use all computation clusters to accelerate one FHE workloads~\cite{f1,craterlake,bts,ark,sharp}, either shallow or deep, \sys utilizes eight cluster affiliations to provide sufficient parallelism for shallow FHE workloads. Worth noting, the computation clusters within a single cluster affiliation share the valuable on-chip L1 cache. As a result, the increased parallelism achieved by adding more swift computation clusters does not come at the expense of significant resource consumption. We will cover more details of this part in \secref{sec:design_cache}.

On the contrary, when it comes to accelerating deep FHE workloads, \sys does not employ the swift clusters. Instead, it utilizes all bootstrappable clusters across cluster affiliations to achieve optimal acceleration for bootstrapping operations. Due to the relatively minor computation resource consumption of the swift clusters, only $<7\%$ hardware resources are not being utilized, as demonstrated in \secref{sec:evaluation_area}.

\parab{Other Modules:} Similar to previous works~\cite{craterlake}, we have implemented real-time key generation and automorphism modules to respectively mitigate off-chip memory bandwidth bottlenecks and accelerate the rotation operations.

\parab{Multi-level Transpose:} In this section, we present our approach of utilizing a multi-level transpose module to establish connectivity between computation clusters both within and across cluster affiliations. The goal of our multi-level transpose is to support both shallow and deep working modes with a varying $N$ from $2^{11}$ to $2^{16}$.

As shown in \figref{fig:design-transpose}, the transpose module in \sys has three levels. When handling shallow FHE workloads, only L1 and L2 transpose modules are used since \sys uses one cluster affiliation to accelerate one shallow task. On the other hand, the L1 and L3 transpose modules are utilized when accelerating a deep workload. We also design a data distributor module to ensure correct data distribution in different working modes.

Our design principle for the transpose module in \sys is to concentrate the design complexity in the L1 transpose module, as it primarily deals with wire placement within a smaller area. In order to mitigate global wiring complexity, we employ fixed and static wiring for implementing the L2 and L3 transpose modules.

The purpose of L1 transpose is to implement local transpose inside each cluster. As we have eight cluster affiliations, the $N$-degree polynomials in each bootstrappable cluster can be regarded as $256 \times D$ matrices, where $D=N/(256*8)$ denotes the input/output length of the NTT pipeline. Similarly, since there are two swift clusters in each cluster affiliations, the polynomials in a swift cluster can be viewed as $128 \times D$ matrices. Inspired by the similarity, we use the $D \times D$ matrix as the basic transpose unit in L1 transpose.

\begin{figure}[t]
\centering
\includegraphics[width=\columnwidth]{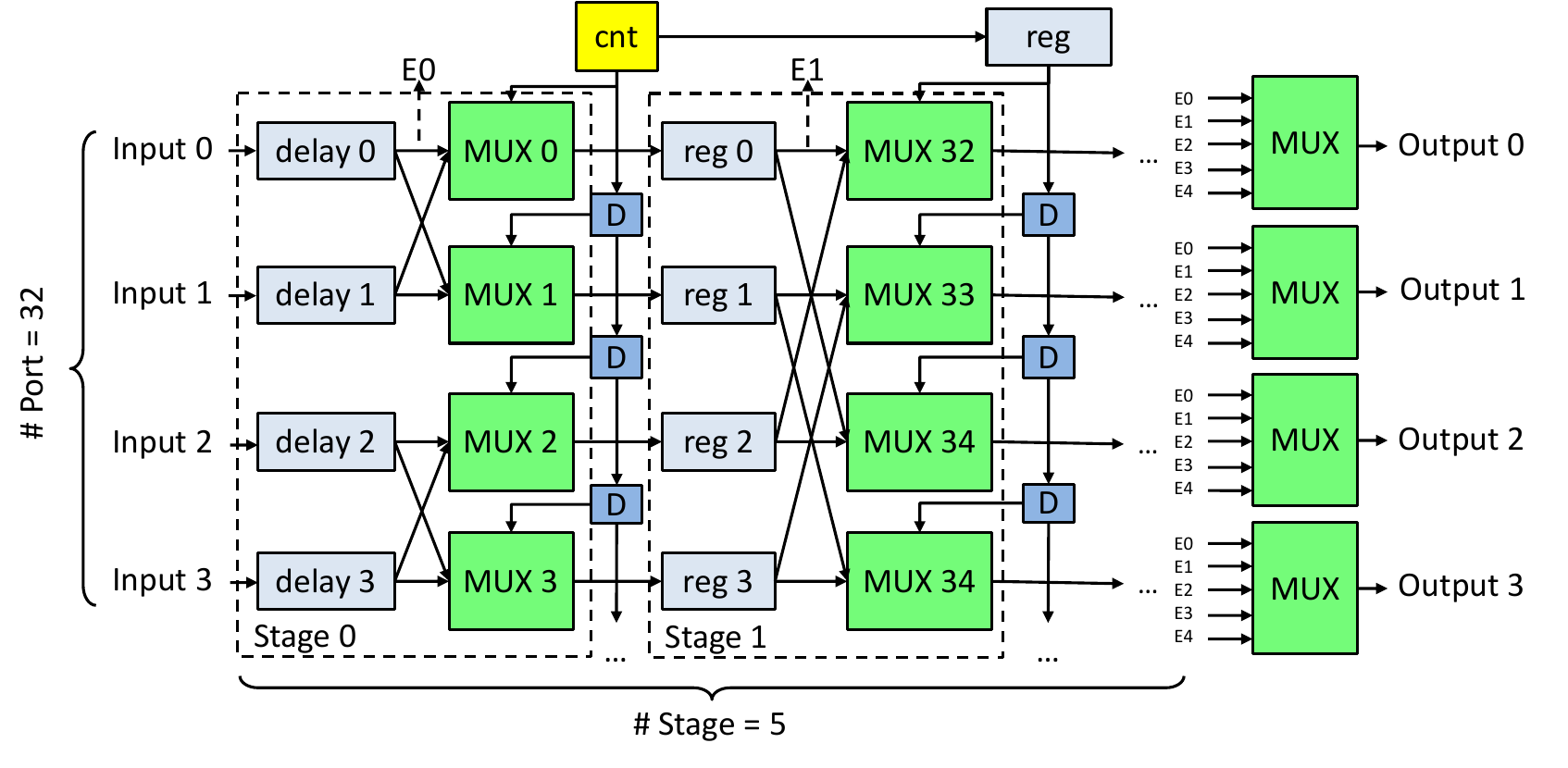}
\caption{Unified building block of L1 transpose module.}
\label{fig:design-transpose-L1}
\end{figure}

The L1 transpose module is built up over multiple building blocks. As shown in \figref{fig:design-transpose-L1}, each of the blocks has 32 input ports and $\rm{log}_2 32$ switching stages, making it possible to concurrently transpose $32/D$ $D \times D$ matrices. In bootstrappable clusters, L1 transpose contains eight building blocks, while in swift clusters, it contains four blocks.

As we support $N\in [2^{11}, 2^{16}]$, $D \in [1, 32]$ is also a varying value. The detailed workflow of our L1 transpose module includes the following steps to dynamically support different $N$ and $D$:

\begin{figure}[t]
\centering
\includegraphics[width=\columnwidth]{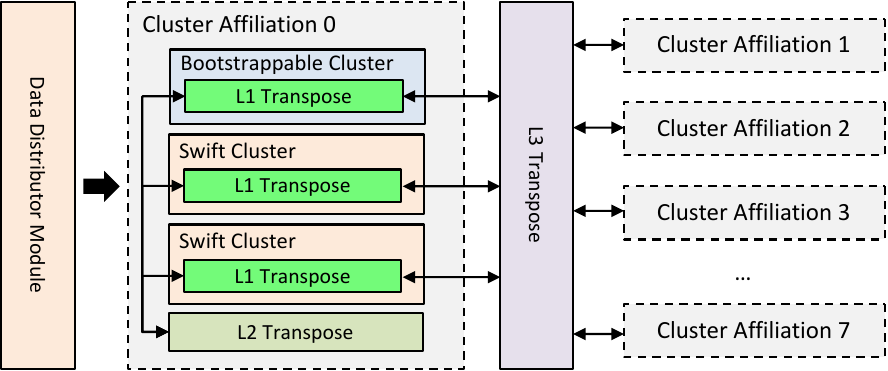}
\caption{Overall design of multi-level transpose.}
\label{fig:design-transpose}
\end{figure}

\begin{icompact}
\item [1.] The delay module delays $i$ cycles for the $i$-th input~(the input data is from L1 cache).
\item [2.] The cnt module is configured as an auto-incremental counter. The MUX module at stage $j$ uses the $j$-th bit~(0 or 1) of the cnt module to determine its data selection. After passing the MUX at stage 0, the L1 transpose building block transposes 16 $2 \times 2$ matrices. Next, after passing the MUX at stage 1, it transposes eight $4\times 4$ matrices. Similar logic applies util the MUX at stage 4 achieves a complete $32 \times 32$ matrix transpose operation.
\item [3.] Since the $D$ is a changing value, we do not always require a complete $32 \times 32$ transpose. We can use the output MUX to select the required output from the pipeline. For example, we can configure the pipeline of the building block to exist at stage 1 to compute a case where $D=4$.
\end{icompact}

The advantages of our design are two-fold: (1) It can achieve a full pipelining, which maximizes the performance. (2) It can achieve high flexibility by only configuring the MUXs, \ie, how to choose data based on bit 0 or 1.

After introducing how L1 transpose module works, we next provide a brief introduction to other modules since they mainly work in a static mode. A detailed workflow and examples of our transpose module are provided in supplemental materials.

The data distribution modules takes a fixed dual-working mode to distribute the data. For shallow workloads, it distributes data within one cluster affiliation~(regard as four clusters with index from 0 to 3), while for deep workloads, it distributes data across all bootstrappable clusters~(eight clusters). Taking an $8192$-point NTT in shallow workloads as an example, based on four-step NTT algorithm, we can regard it as a $128 \times 64$ matrix. Then the data distribution module sends the data column by column~(each column contains 128 rows of data) to four clusters. Specifically, the $i$-th column is sent to L1 cache of the $(i \mod 4)$-th cluster.

The L1-to-L2 transpose connection follows a fixed wiring to achieve a matrix transpose among four clusters. The L2 transpose module has 512 ports in total. The port $i$ of the L1 transpose within cluster $j$ is connected to the $(4 \times i + j)$-th port of L2 transpose.
The L1-to-L3 transpose also follows a fixed connection and the L3 transpose module has 2048 ports in total. Similarly, the port $i$ of the L1 transpose within cluster $j$ is connected to the $(8 \times i + j)$-th port.

\subsection{Scheduler}
\label{sec:design_scheduler}

\sys's scheduler consists of two parts: a software driver that generates control instructions, and a hardware controller that executes these instructions. When it comes to compound operations, such as key-switching, which contains multiple basic operations, \sys's scheduler optimizes their workflows to increase the cache hit ratio. This optimization is achieved purely through hardware implementation. In the following section, we will delve into these details in depth.

The software driver provides standard APIs to integrate \sys with upper-layer FHE libraries like \texttt{Lattigo} and \texttt{SEAL}. It takes key cryptographic parameters, such as $N$ and $L$, as inputs. Similar to previous works~\cite{f1,craterlake,bts,ark,sharp}, the \sys driver generates static control instructions as output. Since FHE workloads are oblivious, meaning that the workflow of an FHE program is independent of its input data, all operations and their dependencies are known in advance. As a result, the \sys software driver can generate instructions that pipeline computation modules in \sys, mitigating most overhead caused by data read/write operations.

However, unlike previous approaches, \sys's instructions explicitly mark the entrances and exits of the corresponding computation cluster's (i)NTT pipeline. The hardware controller follows these instructions to move input data to specific caches, perform computations, and output results to caches or external memory.

The goal of \sys is to provide optimal scheduling policy for both shallow and deep FHE workloads. To achieve it, the scheduler takes the following steps:

\begin{icompact}
\item[1.] \sys analyzes the cryptographic parameters to determine whether the FHE workloads are deep or shallow.
\item[2.] For shallow workloads, the goal of the scheduler is to improve parallelism. Therefore, \sys driver mainly generates instructions to execute parallel small-point (i)NTTs. No data exchange between cluster affiliations is scheduled.
\item[3.] For deep workloads, \sys driver generates instructions of large-point (i)NTTs with BConv operations, which will be executed on all bootstrappable clusters to accelerate the bootstrapping operation as much as possible. We do not use swift computation clusters in deep workloads due to redundant data movement among multiple (i)NTT pipelines.
Currently, we do not use swift computation clusters to execute these instructions,
As we will show in \secref{sec:evaluation_area}, since the swift clusters only consume less than $7\%$ chip area, it does not cause a considerable hardware resource underutilization.
How to efficient combine multiple swift clusters to execute large-point (i)NTTs is one of our future works.
\end{icompact}

\parab{Preemptive Scheduling:} Our scheduler supports preemptive scheduling to efficiently manage mixed workloads. For example, if a deep FHE task is running on \sys and shallow FHE tasks arrive, the deep task can be preempted, allowing the shallow tasks to run first and avoiding the convoy effect. This results in improved average task completion time. At the low level, we provide a priority-based preemptive mechanism, enabling users to assign different priorities to FHE tasks and simulate various scheduling policies, such as shortest-job-first. Preemption is implemented by inserting instructions to temporarily transfer data from SRAM to HBM, loading it back on-demand between tasks.

\subsection{Shared Data Cache}
\label{sec:design_cache}

In this section, we will begin by introducing the hierarchical caching structure implemented in \sys. Then, we will delve into how we use algorithm-level knowledge to determine the appropriate cache volume. %Lastly, we will discuss our approach to mitigate off-chip I/O through instruction preprocessing.

\parab{Hierarchical Caching Structure:}
The hierarchical cache contains one shared L1 cache in each cluster affiliation and one global L2 cache shared by all affiliations.
The L1 caches are designed to facilitate the pipelined execution of individual basic operations within each affiliation, such as NTT or BConv. In \sys, we allocate 8MB SRAM as one L1 cache. In the case of deep workloads, where workloads are distributed among eight affiliations, the L1 cache in each affiliation is capable of supporting polynomial operations with a size of $N \leq 2^{16} / 8$. Moreover, for shallow applications where the size of ciphertexts and keys is significantly smaller, the L1 cache remains sufficient to support one complete shallow workload.

The L2 cache is designed to store data when cluster affiliations need to cooperate with each other in the case of deep FHE workloads. The volume of L2 cache has a significant impact on the overall performance of deep FHE workloads. In our paper, we follow an algorithm-guided approach to decide the volume of L2 cache.

\begin{figure}[t]
\centering
\includegraphics[width=\columnwidth]{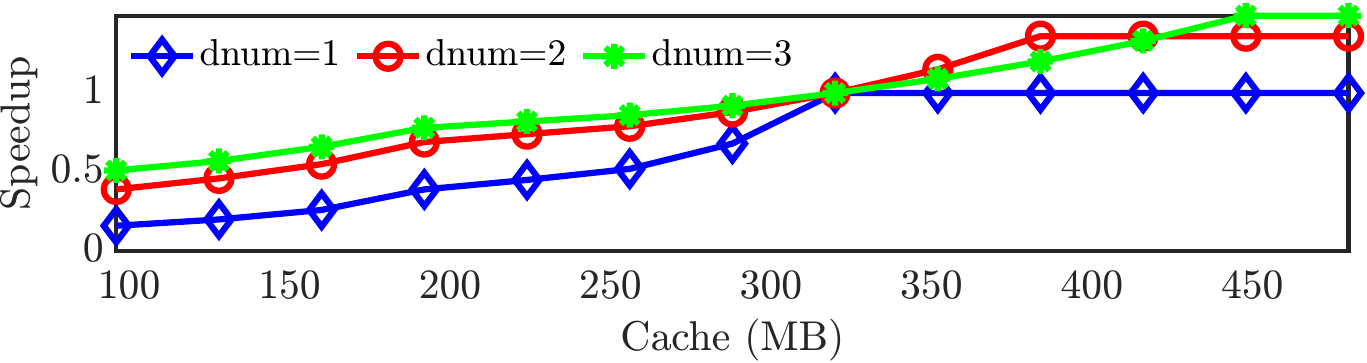}
\caption{Performance of key-switching over varying cache volume.}
\label{fig:design-key-switching-L2}
\end{figure}

After thoroughly analyzing the memory space requirements of operations in deep workflows, we determine that key-switching~(the version used in bootstrapping that requires an iNTT $\rightarrow$ BConv $\rightarrow$ NTT pipeline) is the most time-consuming operation, necessitating a significant amount of memory space to cache keys and intermediate data.
The performance of key-switching with varying total cache volumes is depicted in \figref{fig:design-key-switching-L2}. We evaluate three common $\rm dnum$ settings, namely $\rm dnum=1,2,3$. Our observations indicate that for $\rm dnum=1$, a prevalent setting in most deep learning applications, the optimal performance is achieved when the total cache volume reaches 320MB. Allocating additional L2 cache beyond this point does not yield further performance improvements. Furthermore, 320MB serves as a reasonable setting for $\rm dnum=2,3$, as it does not result in a significant degradation from optimal acceleration. In our paper, \sys targets technology nodes of both 14/12nm and 7nm. Consequently, we have chosen to set the total cache volume to 320MB SRAM, which is a reasonable value for production. However, we suggest that when utilizing a more advanced technology node, such as 3nm, a larger cache could be allocated to further accelerate the key-switching operation for $\rm dnum=2,3$.

The L2 caches are also utilized to store output data that may be reused by subsequent compound operations to eliminate slow off-chip I/O. We use instruction preprocess technology to achieve the goal.

\section{Implementation}
\label{sec:implementation}
We implement \sys in RTL and synthesize it with both an open-sourced 7nm predictive process design kit -- ASAP7~\cite{asap7} -- and a commercial 14/12nm technology library. To achieve a balance between the power consumption and chip area, we choose 1.0 GHz to synthesize most of \sys's components, including \sys's swift and bootstrappable clusters, hardware controller, \etc. The data cache run double pumped at 2.0 GHz, which enables using a single-ported SRAM while serving up to two accesses per clock cycle.

We use HBM2e~\cite{hbm-url} as the off-chip memory. We use two HBM2e PHYs and each can reach a bandwidth of 512 GB/s. Meanwhile, we use the reported data in previous works~\cite{craterlake,bts} to estimate the area and power of the HBM2e memory.

For the software part, we modify the \texttt{Lattigo} library~\cite{lattigo-url} and \texttt{SEAL} library~\cite{seal-url}~(for shallow workloads only) to integrate \sys's functionalities into it, including the software driver, \etc. We also design command-line tools to debug and test \sys. We leverage PCIe 5.0 interface for the communication between \sys and the host since it can offer sufficient bandwidth~\cite{pcie5-url}.
\section{Evaluation}
\label{sec:evaluation}

In this section, we mainly evaluate \sys to demonstrate its effectiveness towards mixed FHE workloads. 
We highlight the following evaluation results.
\begin{icompact}
\item \sys achieves a comparable chip area with other FHE accelerators. The extra area of swift clusters consumes $<7\%$ of the total area~(\secref{sec:evaluation_area}).
\item For deep FHE workloads, \sys achieves an average of $1.4\times$ and $11.2\times$ performance gain than CraterLake and F1+, respectively~(\secref{sec:evaluation_performance}).
\item For shallow FHE workloads, due to \sys's parallel scheduling policy, its performance can achieve up to $8.0\times$ of existing FHE accelerators~(\secref{sec:evaluation_performance}).
\end{icompact}

\subsection{Methodology}
\label{sec:evaluation_methodology}

\parab{Evaluation Method:} \highlight{Since we have not yet taped out \sys and are unable to use an FPGA to implement the entire \sys logic, similar to previous works~\cite{craterlake,bts,f1,ark,sharp}, we evaluate \sys using a cycle-accurate simulator.}

\parab{Compared Schemes:} \highlight{We mainly compare \sys with the following schemes:}

\begin{table}[t]
\small
\centering
\begin{tabularx}{\columnwidth}{b r r}
\toprule
Components & 7nm~(mm$^2$) & 14/12nm~(mm$^2$) \\
\midrule
128-point NTT & 0.50 & 1.42 \\
Modular Mul/Add & 0.31 & 0.91 \\
\multirow{2}*{\shortstack[l]{\textbf{Total. Swift Clusters} \\($16\times$NTT, $16\times$Mod. M/A)}} & \multirow{2}*{12.96} & \multirow{2}*{37.28} \\
\\
\midrule
256-point NTT & 0.99 & 2.81 \\
Modular Mul/Add & 0.63 & 1.81 \\
BConv & 0.69 & 2.01 \\
\multirow{2}*{\shortstack[l]{\textbf{Total. Bootstrappable Clusters} \\($8\times$NTT, $8\times$Mod. M/A, $60\times$BConv)}} & \multirow{2}*{55.09} & \multirow{2}*{160.56} \\
\\
\midrule
Key Generation & 0.73 & 3.00 \\
Automorphism & 3.21 & 9.23 \\
Transpose & 0.13 & 0.37 \\
SRAMs in Clusters & 19.50 & 96.6 \\
Hierarchical Cache & 58.00 & 185.5 \\
$2\times$HBM2e & 29.80 & 29.80 \\
\midrule
\textbf{Total. \sys} & 178.69 & 519.34 \\
\bottomrule
\end{tabularx}
\caption{Area analysis with both 7nm and 14/12nm technology nodes~(Mod. M/A denotes modular multiplication and addition).}
\label{tab:eval-area-analysis}
\end{table}

\begin{icompact}
\item \textbf{F1+} [14/12nm]: F1+ is a upgraded version of F1~\cite{f1}. Similar to the scheme used in \cite{craterlake}, F1+ is scaled to a 256 MB 32-bank scratchpad, 32 compute clusters with 256 lanes each, and 1 MB register file per cluster. However, F1+ suffers from an unoptimized version of key-switching, thus leading to suboptimal performance for deep FHE workloads. The area of F1+ is 636mm$^2$.
\item \textbf{CraterLake} [14/12nm]: CraterLake is the succeeder of F1, which contain 8 256-lane computation group. To compensate the drawback of F1, CraterLake makes all its computing clusters to support bootstrapping by designing an efficient iNTT$\rightarrow$BConv$\rightarrow$NTT pipeline for all of its engines. The chip area of CraterLake is 472mm$^2$.
\item \textbf{ARK} [7nm]: ARK leverages a novel algorithm-architecture co-design to accelerate bootstrapping. The chip area of ARK is 418mm$^2$. %Here, we compare \sys with ARK to report the performance with an advanced technology node, \ie, 7nm.
\item \textbf{SHARP} [7nm]: SHARP utilizes a short word length to reduce the required chip area. The chip area of SHARP is 179mm$^2$.
\item \textbf{Other GPU/FPGA Solutions}: We primarily compare \sys with one GPU-based accelerator, TensorFHE~\cite{tensorfhe}, and two FPGA-based accelerators, HEAX~\cite{heax} and FAB~\cite{fab}.
\end{icompact}

Due to the unavailability of open-source implementations of the FHE accelerators, it is not possible to reproduce their reported performance results independently. Therefore, for the purpose of a fair comparison, we rely on the data provided in the original papers of CraterLake~\cite{craterlake}, ARK~\cite{ark}, and SHARP~\cite{sharp}, similar to previous research works. By utilizing the data from these papers, we aim to establish a meaningful and consistent basis for comparison between \sys and the aforementioned FHE accelerators.

\parab{FHE Applications:} We mainly evaluate \sys with three shallow and four deep FHE applications. Besides the Logistic Regression and LSTM applications mentioned in \secref{sec:analysis}, we further evaluate the following applications:

\begin{figure*}[t]
\begin{minipage}{0.68\textwidth}
\centering
\includegraphics[width=\linewidth]{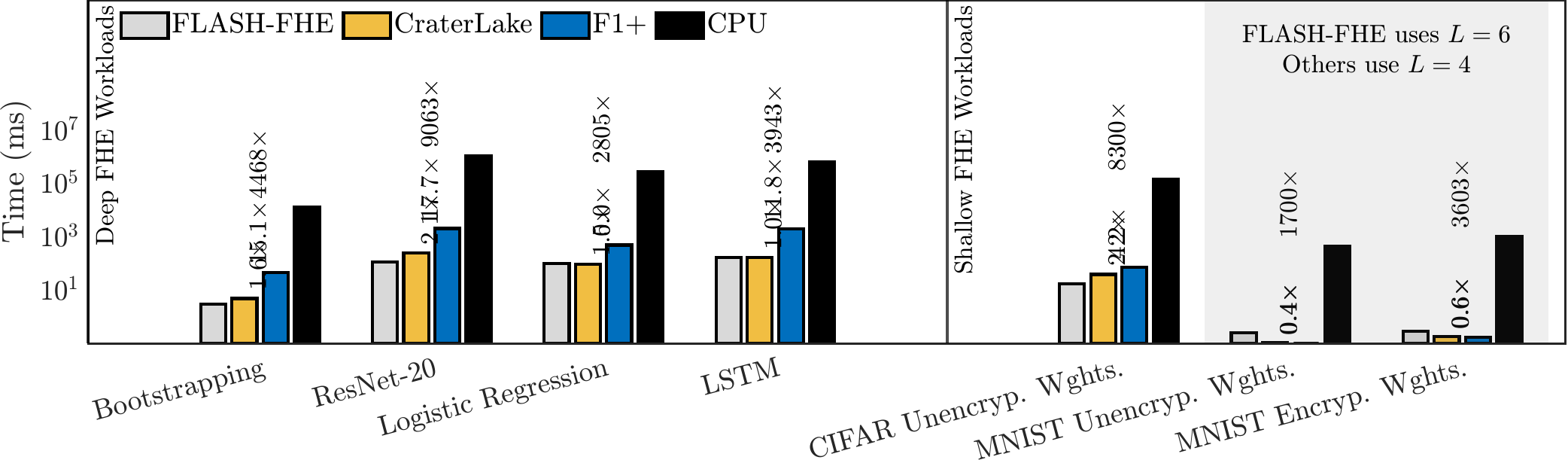}
\caption{[14/12nm Technology Node] Performance of \sys with four deep and three shallow FHE applications.
\label{fig:eval-performance-14nm}
%We compare \sys~(synthesized with the 14/12nm technology node) with CraterLake and F1+, which all use the 14/12nm technology node. We also compare \sys with an AMD Ryzen Threadripper PRO 3975WX 32-core, 64-thread CPU. The batch size of Logistic Regression is 256.
}
\end{minipage}
\hfill
\begin{minipage}{0.29\textwidth}
\centering
\includegraphics[width=\linewidth]{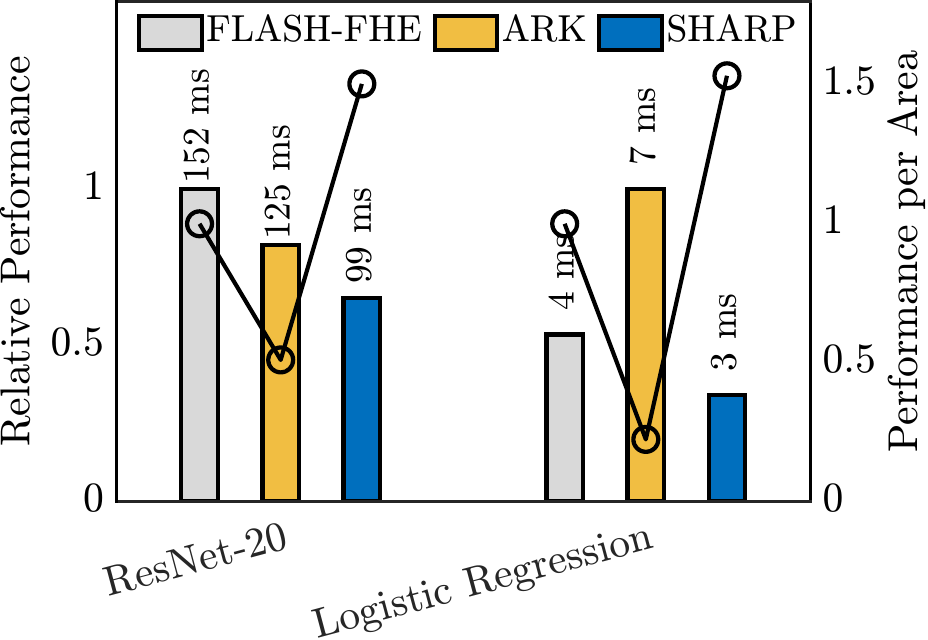}
\caption{[7nm Technology Node] Performance of \sys.}
\label{fig:eval-performance-7nm}
\end{minipage}
\end{figure*}

\begin{icompact}
\item \textbf{LoLa-CIFAR with Unencrypted Weights:} LoLa is a low latency privacy-preserving neural network~\cite{lola}. LoLa contains three different variants and all of them are shallow FHE workloads. The first variant is with the CIFAR-10 dataset~\cite{cipher-10}. The parameters of the neural network are not encrypted. We set $N=2^{13}$ and $L=7$.
\item \textbf{LoLa-MNIST with Unencrypted Weights:} This is the second variant of LoLa with MNIST dataset~\cite{mnist}. The parameters of the neural network are not encrypted. We choose $N=2^{13}$ and $L=6$.
\item \textbf{LoLa-MNIST with Encrypted Weights:} This is the third variant of LoLa with MNIST dataset, which further adopts encrypted weights for the neural network. We set $N=2^{13}$ and $L=6$ for this application.
\item \textbf{Packed Bootstrapping:} Packed bootstrapping is a deep workload. Similar to \cite{craterlake}, the packed bootstrapping takes a ciphertext with $N=2^{16}$ and $L=3$ as the initial parameters, and then exhausts its multiplicative level by bringing $L$ to 57. Afterwards, it performs bootstrapping to refresh the multiplication level of the ciphertext. In packed bootstrapping, different from unpacked bootstrapping, the ciphertext uses all available slots, \ie, $N/2$. The packed bootstrapping is also the bootstrapping algorithm in other deep workloads. We set $N=2^{16}$ and $L=57$.
\item \textbf{ResNet-20:} It's a deep FHE workload performing neural network inference with ResNet-20~\cite{resnet}, whose parameters are encrypted via CKKS. We mainly refer the implementation mentioned in \cite{resnet-ckks}. We set $N=2^{16}$ and $L=41$.
\end{icompact}

For LSTM, we set $N=2^{16}$ and $L=13$. For the Logistic Regression evaluation, we will utilize batch sizes of both 256 and 1024, with a feature size of 256. We set $N=2^{13}$ and $L=33$. These settings are chosen to align with the evaluation parameters used in CraterLake and ARK.

\subsection{Area Analysis}
\label{sec:evaluation_area}
In this section, we present the area analysis for \sys. Following the approach of previous works such as BTS~\cite{bts}, SHARP~\cite{sharp}, and ARK~\cite{ark}, we employ FinCACTI~\cite{fincacti} to model the SRAMs and caches utilized in \sys. The results of the analysis are summarized in \tabref{tab:eval-area-analysis}.

With the 14/12nm technology node, \sys occupies a total area of 519.34mm$^2$, which is comparable to that of CraterLake~\cite{craterlake}. Notably, \sys incorporates 16 swift engines in its design, significantly enhancing the performance of shallow FHE workloads, as demonstrated in \secref{sec:evaluation_performance}. 

Furthermore, the large performance gain is achieved only with the addition of a relatively small portion of the chip area~(the logic circuits of swift clusters occupy less than $7\%$ of the total area) since swift clusters share the memory resources of bootstrappable clusters.

As for other components, the logic circuits of \sys's bootstrappable computation clusters account for $31.1\%$ of the area, while the total cache occupies $55.6\%$. Since \sys allots 320MB of cache to enhance the performance of key-switching operation~(refer \secref{sec:design_cache} for more details), the SRAMs consume slightly more chip area compared to other FHE accelerators. For the 7nm technology node, the total chip area of \sys is 178.69mm$^2$, which aligns with the estimated scaling from 14/12nm to 7nm area~(approximately $2.9\times$)\cite{scaling}.

\subsection{Performance}
\label{sec:evaluation_performance}

In this section, we compare the performance of \sys with other typical FHE accelerators. First, we compare \sys synthesized with the 14/12nm technology node against other FHE accelerators using the same technology, namely CraterLake and F1+. Subsequently, we compare \sys implemented with the 7nm technology node with ARK and SHARP, which is also based on the 7nm technology. Additionally, we compare \sys with an AMD Ryzen Threadripper PRO 3975WX CPU, featuring 32 cores and 64 threads.

\sys and CraterLake employ 128-bit security for all deep FHE workloads and 80-bit security for shallow FHE workloads, which we consider practical for production environments. F1+ employs 80-bit security for all seven workloads. ARK and SHARP use 128-bit security for their two deep FHE workloads.

\parab{Single Workload:} We first evaluate all accelerators with one single FHE workload at a time and the performance of \sys at the 14/12nm technology node is presented in \figref{fig:eval-performance-14nm}. Here, we highlight the following results. 
First, across all four deep workloads, \sys can achieve an average~(geometric mean) speedup of $1.4\times$ and $11.2\times$ compared to CraterLake and F1+, respectively. The performance improvement over CraterLake is mainly due to the adequate cache~(320MB in \sys v.s. 256MB in CraterLake), which can boost the performance of key-switching operation~(as shown in \secref{sec:design_cache}).

Second, for the three shallow FHE workloads, \sys can achieve a speedup of $0.4 - 2.2\times$ and $0.4 - 4.2\times$ compared to CraterLake and F1, respectively. For the specific scenarios of LoLa-MNIST with and without Unencrypted Weights, it is worth noting that \sys employs a value of $L=6$, while F1 and CraterLake utilize $L=4$ since we encounter difficulties in using $L=4$ to launch the two applications. Consequently, it is important to acknowledge that \sys's performance in this context is inferior to that of CraterLake and F1+. However, in the subsequent sections, we will show that even with a larger $L$ setting, \sys can achieve significantly higher performance when there are multiple shallow FHE workloads.

Next, we compare \sys implemented with the 7nm technology node against ARK and SHARP. Since ARK and SHARP do not provide performance results specifically for shallow FHE workloads, we focus our comparison on two typical deep workloads: ResNet-20 and Logistic Regression. Since some accelerators with the 7nm technology utilize dramatic chip area, we also show the metric of performance per area in this part.

The evaluation results are presented in \figref{fig:eval-performance-7nm}, and we highlight the following points.
Compared to ARK, \sys achieves $42.3\%$ better performance in Logistic Regression but $21.6$ worse performance in ResNet-20. However, \sys achieves consistency higher performance per area, from $1.49\times$ to $1.78\times$. The reason is that although ARK adopts an algorithm hardware co-design method to optimize the performance, it results in a dramatically larger chip area, leading to much worse performance per chip area. SHARP achieves better absolute performance and performance per chip area than \sys since it adopts a short-word size optimization method, which is not utilized in the current implementation of \sys. Worth noting, \sys's idea of leveraging a heterogeneous architecture is \emph{orthogonal} to the optimizations used in ARK and SHARP. Therefore, we can combine \sys with them to further improve their performance for mixed FHE workloads.

\parab{Multiple Shallow Workloads:} In this part, we evaluate \sys's performance~(implemented with 14/12nm technology node) when there are multiple shallow workloads. We vary the number of shallow workloads from one to ten and measure the average execution time. The results are shown in \figref{fig:performance-shallow-workloads}. We observe when there are more than three shallow workloads, \sys can achieve a much better average execution time than CraterLake even when \sys uses $L=6$. The reason is that \sys can execute multiple shallow workloads in parallel while CraterLake can only use a sequential scheduling policy due to its homogeneous design. \sys can achieve up to $8.0\times$ speedup when all of its cluster affiliations are utilized.

\begin{figure}[t]
\centering
\includegraphics[width=\columnwidth]{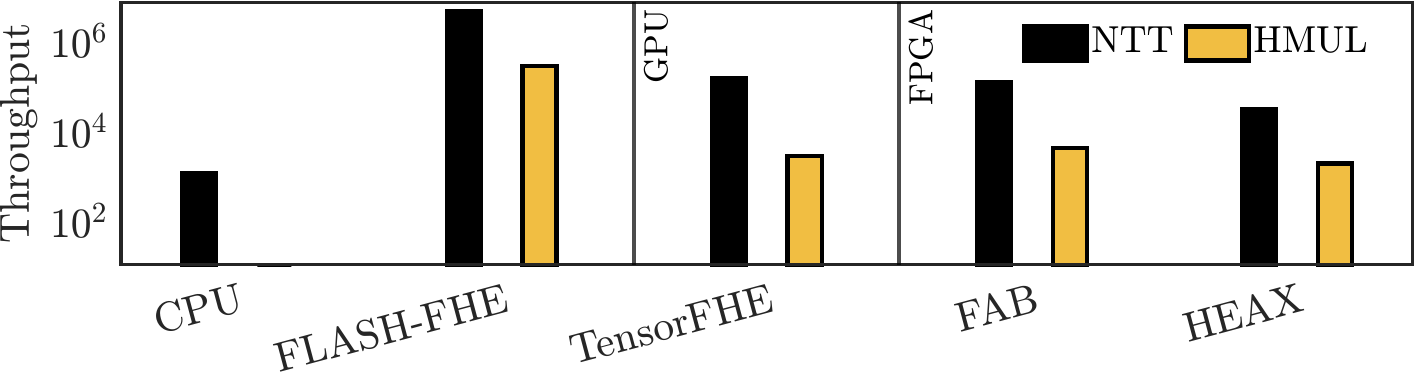}
\caption{Performance of \sys compared to representative GPU and FPGA solutions.}
\label{fig:performance-gpu-fpga}
\end{figure}

\parab{Comparison with GPU/FPGA:} In this section, we highlight \sys's advantages over GPU/FPGA solutions for shallow workloads. Due to the lack of data on shallow workloads in previous works, we instead compare the performance of NTT and Homomorphic Multiplication (HMUL), which are the primary computations in shallow workloads and account for the majority of execution time. For a fair comparison, the evaluation is conducted using the same shallow parameters as in previous works~(\ie, $N=2^{14}$, $\log PQ=438$). As shown in \figref{fig:performance-gpu-fpga} (note the Y-axis is in log scale), \sys achieves more than $30\times$ throughput~(measured in number of executed operations) compared to the state-of-the-art GPU-based solution, TensorFHE~\cite{tensorfhe}, and FPGA-based designs, FAB~\cite{fab} and HEAX~\cite{heax}, for NTT computation. For HMUL operations, \sys provides a $60-100\times$ acceleration compared to the FPGA and GPU implementations. These results demonstrate the significant performance improvement of \sys over existing GPU/FPGA solutions for shallow workloads.

\subsection{Power Consumption}
\label{sec:evaluation_power}

\begin{figure}[t]
\begin{minipage}{0.65\columnwidth}
\includegraphics[width=\textwidth]{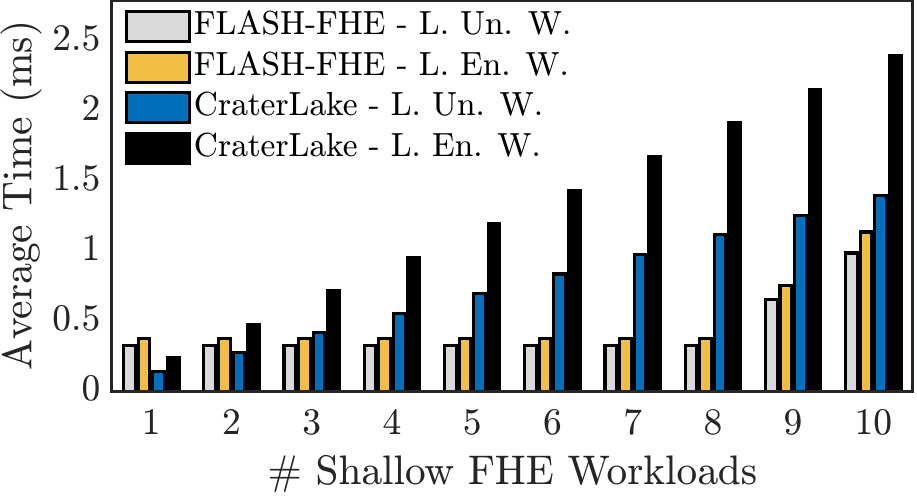}
\caption{[14/12nm Technology Node] Performance of \sys with parallel shallow FHE workloads.}
\label{fig:performance-shallow-workloads}
\end{minipage}
\hfill
\begin{minipage}{0.3\columnwidth}
\includegraphics[width=\textwidth]{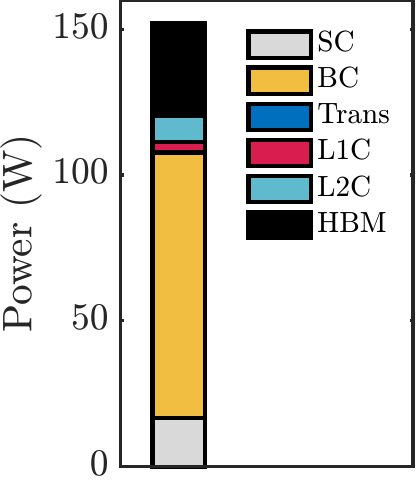}
\caption{Power consumption breakdown.}
\label{fig:power-breakdown}
\end{minipage}
\end{figure}

We also conducted an evaluation of the power consumption of \sys using the 14/12nm technology node, as we currently lack access to the power budget for certain logic modules in the 7nm technology. The total power consumption of \sys amounts to 152.11W, which is superior to CraterLake~(317W) and ARK~(281.3W), and comparable to BTS~(163.2W). \figref{fig:power-breakdown} provides a breakdown of \sys's power consumption and shows the power consumption of the bootstrappable clusters~(BC), swift clusters~(SC), transpose unit~(Trans), L1 cache~(L1C), L2 cache~(L2C) ad HBM. The results demonstrate that the bootstrappable computation cluster accounts for the majority of power consumption, reaching up to $60.0\%$, while the swift clusters consume only $11.0\%$ of the total power. 
Therefore, our results further indicate that the improved performance resulting from adding more swift clusters does not significantly increase power consumption.

\section{Discussion}
\label{sec:discussion}
\parab{Integration with Other FHE Accelerators:} As discussed, it has become evident that several FHE accelerators employ dedicated optimization methods to achieve further performance gain~\cite{ark,sharp}. One such example is SHARP~\cite{sharp}, which utilizes a short word length to minimize chip area, albeit with a slight accuracy trade-off. 
%On the other hand, ARK~\cite{ark} enhances its performance by co-designing its bootstrapping algorithm and hardware architecture. 
In our paper, we propose \sys as an alternative approach to achieve high performance, employing a heterogeneous architecture tailored for mixed FHE workloads. Since \sys's idea is orthogonal to SHARP and ARK, we believe that integrating \sys with these existing FHE accelerators can yield superior performance outcomes.

\parab{Combining Multiple Swift Clusters for Deep Workloads:} In the current design of \sys, we have focused on decomposing a bootstrappable computation cluster to support multiple shallow FHE workloads. However, we acknowledge that our system does not currently provide support for combining multiple swift computation clusters to accelerate a single deep FHE workload. This limitation stems from the challenges associated with complex computation scheduling among multiple (i)NTT pipelines. Such scheduling inherently involves massive data movement, which can potentially degrade performance. %Given that the resource underutilization resulting from the absence of swift engines for deep FHE workloads is deemed acceptable, 
Addressing this issue has been left as a future research direction for \sys.
\section{Related Works}
\label{sec:related_works}
Besides the related works discussed in \secref{sec:background_fhe_acc}, we further cover the following two more directions in this section.

\parab{Other Optimizations for FHE:} In addition to leveraging accelerators for enhanced performance in fully homomorphic encryption (FHE), there is an alternative approach focused on algorithm-level optimization. 
While the accelerators discussed in \secref{sec:background_fhe_acc} primarily target second-generation FHE schemes such as BFV~\cite{bfv}, BGV~\cite{bgv}, and CKKS~\cite{ckks}, third-generation FHE schemes based on the GSW scheme~\cite{gsw} have been proposed, such as TFHE~\cite{tfhe}. These schemes uses exceptionally fast bootstrapping, often completing in less than 0.1 seconds. However, they come with the drawback of incompatibilities with batching, introducing new trade-offs that limit their usability. Strix introduces the first FHE accelerator for TFHE~\cite{strix}, where it designs a two-level ciphertext batching method with programmable bootstrapping. Another direction of exploration involves optimizing software implementations. Researchers have developed efficient software libraries, including \texttt{SEAL}~\cite{seal-url} and \texttt{Lattigo}~\cite{lattigo-url}, or compilers such as EVA~\cite{eva}, to improve FHE performance. %Furthermore, FHE compilers like EVA~\cite{eva} have been introduced to automatically generate efficient implementations of FHE workloads based on these libraries, aiming to further enhance the usability.

\parab{Partial Homomorphic Encryption Acceleration:} Several existing works focus on enhancing the performance of partial homomorphic encryption~(PHE) schemes, especially the Paillier scheme. Notably, FLASH employs FPGA technology to expedite PHE operations specifically for federated learning~\cite{flash}. Similarly, HAFLO utilizes GPUs to accomplish a comparable objective~\cite{haflo}. Additionally, Shi \etal have developed a dedicated 28nm ASIC tailored for accelerating PHE processes~\cite{paillier_chip}. Since the basic building blocks of the Paillier scheme are modular multiplications and exponentiations, the design choices of these PHE accelerators are very different from FHE accelerators, which are built on polynomials computations.
\section{Conclusion}
This paper proposed \sys, the first heterogeneous acceleration architecture for FHE. We have provided a full RTL implementation of \sys and synthesized it using two representative technology nodes. Experiments with three shallow and four deep FHE workloads show that \sys can consistently achieve high performance for mixed FHE workloads.

\newpage

%-------------------------------------------------------------------------------
%\clearpage
\bibliographystyle{plain}
\bibliography{reference}

\end{document}